%% file: paper.tex
\documentclass[10pt,conference]{IEEEtran}

\usepackage{tcolorbox}
\usepackage{booktabs}
\usepackage[sort&compress, numbers]{natbib}
\usepackage{graphicx}
\usepackage{booktabs}
\usepackage{relsize}
\usepackage{colortbl}
\usepackage{enumerate}
\usepackage{enumitem}
\setlist{leftmargin=5.5mm}
\usepackage{amsmath}
\usepackage{tikz}
\usepackage{pgf}
\usepackage{pbox}
\usepackage{rotating}
\usepackage{multirow}
\usepackage{balance}
\usepackage{subfigure}
\usepackage[english]{babel}
\usepackage{soul}
\usepackage{url}

\usepackage{extarrows}

\newcommand{\startlist}{\begin{list}{\labelitemi}{\leftmargin=1em}\setlength{\itemsep}{-1mm}}
\newcommand{\stoplist}{\end{list}}

\newcommand{\smallsection}[1]{\underline{\bf #1}.}
\newcommand{\ea}{\textit{et al.}}

\newcolumntype{?}{!{\vrule width 2pt}}

\usepackage{arydshln} 

\newcommand{\cnter}{developer}
\newcommand{\analysisI}{The differences of code ownership approximations}
\newcommand{\analysisII}{The association of code ownership approximations with software quality}

\newcommand{\rqone}{To what degree do sets of developers identified by the commit-based and line-based approaches differ?}
\newcommand{\rqtwo}{Do the commit-based and line-based approximations provide consistent estimates of ownership values?}
\newcommand{\rqthree}{To what degree are major developers missed by the commit-based and line-based approximations?}

\newcommand{\rqfour}{Which code ownership approximations have the strongest association with software quality?}

\newcommand{\rqfive}{Whether code ownership that is important in the past release is also important in the current release?}

\begin{document}
\title{Code Ownership: The Principles, Differences, and Their Associations with Software Quality}



\author{
    \IEEEauthorblockN{Patanamon Thongtanunam}
    \IEEEauthorblockA{
        \textit{The University of Melbourne, Australia.}\\
        patanamon.t@unimelb.edu.au
    }
\and
    \IEEEauthorblockN{Chakkrit Tantithamthavorn}
    \IEEEauthorblockA{ 
        \textit{Monash University, Australia.}\\
        chakkrit@monash.edu
    }
}

        


\IEEEtitleabstractindextext{%
\begin{abstract}
Code ownership--an approximation of the degree of ownership of a software component---is one of the important software measures used in quality improvement plans.
However, prior studies proposed different variants of code ownership approximations.
Yet, little is known about the difference in code ownership approximations and their association with software quality.
In this paper, we investigate the differences in the commonly used ownership approximations (i.e., commit-based and line-based) in terms of the set of developers, the approximated code ownership values, and the expertise level.
Then, we analyze the association of each code ownership approximation with the defect-proneness.
Through an empirical study of 25 releases that span real-world open-source software systems, we find that commit-based and line-based ownership approximations produce different sets of developers, different code ownership values, and different sets of major developers.
In addition, we find that the commit-based approximation has a stronger association with software quality than the line-based approximation.
Based on our analysis, we recommend line-based code ownership be used for accountability purposes (e.g., authorship attribution, intellectual property), while commit-based code ownership should be used for rapid bug-fixing and charting quality improvement plans.
\end{abstract}

\begin{IEEEkeywords}
Code Ownership Approximation, Software Quality Assurance.
\end{IEEEkeywords}}

\maketitle

\pagestyle{plain}

\IEEEdisplaynontitleabstractindextext
\IEEEpeerreviewmaketitle

\input{sections/introduction.tex}

\input{sections/background.tex}
\input{sections/part-one.tex}
\input{sections/part-two.tex}

\section{Discussion and Implications}\label{sec:implication}


Our results have highlighted the key differences in approximations between the commonly used code ownership approximations, i.e., commit-based and line-based approaches.
Specifically, our RQ1-RQ3 shows that the set of developers and their ownership values identified by the two approaches are different and our RQ4-RQ5 show the important scores among ownership metrics are different.
These results highlight that these code ownership approximations capture different aspects of the contributions and relationship with software quality.
Thus, a choice of code ownership approximation should be based on the key goal of code ownership usage.

\textbf{Recommendation 1.} The line-based approach would be more suitable for accountability (e.g., authorship attribution, intellectual property) since our RQ1 also shows that the line-based approach tends to identify a larger set of developers than the commit-based approach.
However, our RQ4 and RQ5 show that the number of code authors and the ownership values approximated by the line-based ownership should not be a concern for a quality improvement plan as these metrics are weakly associated with defect-proneness.

\textbf{Recommendation 2.} The commit-based approach would be more suitable for rapid bug-fixing and quality improvement plans.
RQ3 shows that a large proportion of developers who were missed by one of the approaches are major developers who made a large contribution.
Developers who have recently made changes to the file are likely to possess a more up-to-date understanding of the code compared to their counterparts who have not been actively involved in the recent development cycle.
The results of RQ4 and RQ5 confirm that the commit-based ownership metric is the most important metric when explaining the defect model trained from past releases and also when explaining defective files in the current releases.
For charting a quality improvement plan, practitioners may use a LIME's explanation which provides the association rules between code ownership metrics and defect-proneness.
For example, Given an association rule of \{$\mathrm{COMMIT\_OWN}<$~0.5\}$~\Rightarrow$~BUG, which suggests that \emph{if the commit-based ownership metric is less than 0.5, there is a higher chance that a file will be defective}.
Then, by flipping the LIME association rule, \{$\mathrm{COMMIT\_OWN}>=$~0.5\}$~\Rightarrow$~CLEAN, a quality improvement plan can be as maintaining a degree of commit-based ownership higher than 0.5 to reduce the risk of having defects.

\input{sections/relatedwork.tex}

\section{Threats to the Validity}\label{sec:threats}

We now discuss the threats to the validity of our study.

\smallsection{Construct Validity} 
In addition to heuristic-based code ownership approximations studied in this work,  machine learning (ML) or statistical approaches can also be used~\cite{Cury2022Identifying}. 
However, the key goal of this paper is to empirically demonstrate the impact of different aspects captured by different code ownership approximations on the ownership values and identified owners.
We believe that the key findings of our paper remain unchanged when considering other code ownership approximation techniques.

\smallsection{Internal Validity}
We studied a limited number of confounding factors for our case study, as there might be other confounding factors that are associated with software quality.
To mitigate this threat, we chose to study a large number of software metrics (i.e., 64 metrics) that capture both process and product dimensions.
Thus, other confounding factors can be included in future work.

\smallsection{External Validity}
We studied a limited number of software systems.
Thus, our results may not generalize to other datasets, domains, or ecosystems. 
However, we mitigated this threat by choosing a range of different non-trivial, real-world, open-source software applications.
Nonetheless, additional replication studies in a proprietary setting and other ecosystems will prove useful.

\section{Conclusions}\label{sec:conclusion}

In this paper, we find that commit-based and line-based ownership approximations produce different sets of developers, different code ownership values, and different sets of major developers. In addition, we find that the commit-based approximation has a stronger association with software quality than the line-based approximation, suggesting that commit-based code ownership approximations should be used to accurately explain why a file is predicted as defective and to guide the development of QA improvement plans.
Based on our analysis, we recommend line-based code ownership be used for accountability purposes (e.g., authorship attribution, intellectual property), while commit-based code ownership should be used for rapid bug-fixing and charting quality improvement plans.

\balance

\small
\bibliographystyle{ACM-Reference-Format}
\bibliography{filteredref}

\end{document}

%% file: sections/introduction.tex
\section{Introduction}

Code ownership is used to establish a chain of responsibility for software artifacts.
In large-scale software organizations, a large number of software components are developed.
Thus, code ownership~\cite{ahlgren2020ownership,kalgutkar2019code} is highly important for software-intensive organizations (e.g., Facebook~\cite{ahlgren2020ownership}, Microsoft~\cite{Bird2011a,bird2010analysis}) to identify a responsible (or suitable) developer of a given piece of source code as well as to attribute copyrights and Intellectual Properties (IP) to the right owner.
Prior studies showed that code ownership can be used to recommend developers who should fix a bug~\cite{ahlgren2020ownership,anvik2006should,tantithamthavorn2015impact} or who should review a patch~\cite{sadowski2018modern,Al-Zubaidi2020}.
In addition, prior studies also suggested that code ownership should be considered in a quality management plan to mitigate the risk of having software defects~\cite{thongtanunam2016revisiting,Bird2011a,bird2010analysis}. 



At the heart of code ownership management is the operationalization (approximations) of code ownership.
Prior studies have proposed variants of code ownership approximations~\cite{Bird2011a,bird2010analysis,rahman2013and}.
Two conventional approximations are (1) the commit-based approach of Bird~\ea~\cite{Bird2011a,bird2010analysis} and (2) the line-based approach of Rahman and Devanbu~\cite{rahman2013and}.
These two code ownership approximations are often used in empirical studies related to software quality~\cite{yatish2019mining, ahlgren2020ownership, Bird2011a,bird2010analysis}. 
The underlying intuition for the commit-based approach is \textit{the more frequent the code changes (i.e., commits) made by a developer to a file, the higher ownership value the developer should have}~\cite{Bird2011a,bird2010analysis}.
On the other hand, the intuition for the line-based approach is \textit{the larger the proportion of code lines in a file authored by a developer, the higher ownership value the developer should have}~\cite{rahman2013and}. 


Since the commit-based and line-based code ownership approximations use a different granularity of information (i.e., commits or files), they might produce a different set of owners.
For example, a developer who contributes a large proportion of code in one change may be considered as a major \cnter{} when using the line-based approximation but may be considered as a minor \cnter{} when using the commit-based approximation.
However, there exist no studies investigating the difference in code ownership approximations.
Furthermore, prior work also found that code ownership values are associated with defect-proneness~\cite{Bird2011a,bird2010analysis,rahman2013and}.
However, there exist no studies investigating which code ownership approximations should be considered when developing defect prediction models for charting software quality improvement plans.
Thus, a lack of understanding of the difference in code ownership approximations could lead to incorrect ownership attribution, misleading owner identification, and suboptimal quality improvement plans.

In this paper, we investigate the differences in code ownership approximations along three dimensions, i.e., the set of developers, the approximated code ownership values, and the expertise level, and investigate their associations with software quality.
Through an empirical study of 25 releases that span seven large-scale software systems, we address the following research questions.

\begin{enumerate}[label=\textbf{RQ\arabic*},leftmargin=*]
    \item \textbf{\rqone}\\
    \underline{Summary:} Only 0\% to 40\% of developers can be identified by both commit-based and line-based approaches, indicating these approaches do not identify the same set of developers. A substantial proportion of developers can be missed when using one of the ownership approximation approaches. 
    \item \textbf{\rqtwo}\\
    \underline{Summary:} For the developers who can be commonly identified by both commit-based and line-based approaches, their ownership values are different and not highly consistent. Nevertheless, a majority (50\% to 93\%) of these developers still have a consistent expertise level in the two approaches.
    \item \textbf{\rqthree}\\
    \underline{Summary:} The developers who are missed by the line-based approach have an ownership value that is statistically higher than the developers who are missed by the commit-based approach.
    Moreover, a large proportion of these developers are major developers, suggesting that major developers can be missed when using one of the ownership approximation approaches.
    \item \textbf{\rqfour}\\
    \underline{Summary:} Considering all ownership metrics and confounding factors, the commit-based code ownership metrics have the highest important scores to our software defect models, suggesting the strongest relationship of these metrics with software quality.
    \item \textbf{\rqfive}\\
    \underline{Summary:} The commit-based code ownership which is important in the past release is also important in the current release, suggesting that commit-based code ownership should be used to explain why a file is predicted as defective and guide the development of QA improvement plans.
\end{enumerate}


Our results lead us to conclude that commit-based and line-based code ownership approximations identify almost different sets of developers, weakly inconsistent ownership values, and could have missed some major developers, confirming that care should be taken when selecting a code ownership approximation.
Based on our analysis, we recommend line-based code ownership be used for accountability purposes (e.g., authorship attribution, intellectual property), while commit-based code ownership should be used for rapid bug-fixing and charting quality improvement plans.


\textbf{Novelty \& Contributions:} We are the first to present:
\begin{itemize}
    \item The degree of the differences between the two commonly-used code ownership approximations (i.e., commit-based and line-based) in three main aspects, i.e., identified developers, ownership values, and expertise levels (RQ1, RQ2, RQ3).
    \item An investigation of the association with software quality between ownership metrics approximated by the commit-based and line-based approaches (RQ4, RQ5).
    \item A recommendation on when to use commit-based and line-based code ownership approximations.
    \item The replication package is available to facilitate in GitHub.\footnote{https://github.com/awsm-research/code-ownership}
\end{itemize}

\textbf{Organization:} 
Section~\ref{sec:background} provides background and motivating scenarios.
Section~\ref{sec:analysisone} presents the investigation of the differences in code ownership approximations.
Section~\ref{sec:analysistwo} presents the association of code ownership approximations and software quality.
Section~\ref{sec:implication} discusses the implications of our results and provides recommendations.
Section~\ref{sec:relatedwork} discusses the related work, while Section~\ref{sec:threats} discusses the threats to the validity.
Section~\ref{sec:conclusion} concludes the paper.

%% file: sections/background.tex
\begin{figure*}[t]
\centering
\includegraphics[width=.8\textwidth]{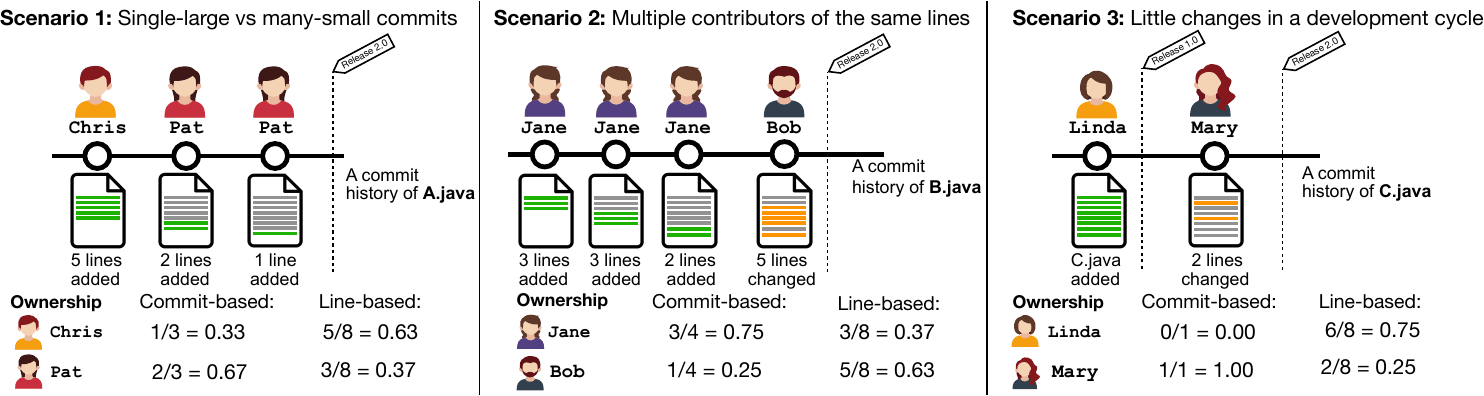}
\caption{Illustrative scenarios where the commit-based and line-based code ownership approaches will approximate different ownership values.}
    \label{fig:motivating}
\end{figure*}

\section{The Principles of Code Ownership and Their Approximations}\label{sec:background}

\subsection{Code ownership}
Code ownership refers to the concept of establishing a chain of responsibility for specific parts of the codebase to individual developers or teams.
These measures can help managers and stakeholders understand which developers or teams are most knowledgeable about a codebase, who is contributing the most to its development, and who is best equipped to maintain and improve it over time.
Code ownership can lead to a more efficient, collaborative, and higher-quality development process.
Specifically, the benefits of code ownership include:

\begin{itemize}

    \item \textbf{Accountability~\cite{kothari2007probabilistic,liu2021practical,nordberg2003managing,oliveira2020code}.} Code ownership creates accountability which helps the team identify who wrote (or is responsible for) this part of code. Further, this helps the team to establish authorship attribution and quantify the individual productivity of developers.
    Other team members are encouraged to communicate with each other about dependencies and potential conflicts. This can help prevent integration issues and improve collaboration across the team.

    \item\textbf{Rapid bug fixing~\cite{anvik2006should,linares2012triaging,borg2023u}.} With code ownership, developers are intimately familiar with the code they wrote or are responsible for, making it easier and faster for them to fix bugs that arise. This can lead to more efficient bug resolution and quicker turnaround times.

    \item\textbf{Increased code quality~\cite{rahman2011ownership,rahman2011ownership,greiler2015code,bass2015product}.} When developers take ownership of their code, they gain familiarity and deep domain expertise in the areas that they are responsible for. This can result in high-quality design and coding. 



\end{itemize}


\subsection{Code ownership approximation} 
Code ownership approximation refers to the process of measuring the degree of code ownership that an individual developer or team has over a codebase or software project. 
Various data were used to approximate code ownership such as code commits~\cite{bird2009does,Bird2009}, lines of code~\cite{ullah2019source,bogomolov2021authorship,meng2013mining}, bug reports~\cite{Banitaan2013}, code reviews~\cite{thongtanunam2016revisiting}, time~\cite{daSilva2015, Robbes2013}, organizational structure~\cite{nagappan2008influence}, and social coding platform like GitHub~\cite{Constantinou2016}. 
A recent mapping study reports that majority of code ownership approximation approaches are based on measuring the number of code commits and lines of code~\cite{Brasil-Silva2022}.
Thus, in this work, we focus on these two common code ownership approximation approaches:


\begin{itemize}
\item \textbf{Commit-based Code Ownership.} This measure calculates the number of code changes made by a particular developer or team over a certain period of time, such as a week or a month.
The code change ownership is used by Bird~\ea~\cite{Bird2011a,bird2010analysis} to measure the proportion of commits that a developer has contributed to a file relative to the total number of commits for that file.
Formally, the commit-based code ownership of a given developer $d$ for a file $f$ is computed as follows:

{\small
\begin{equation}
    \mathrm{OWN\_COMMIT}(d,f) = \frac{\mathrm{\#commits}(d,f)}{\mathrm{\#total\_commits}(f)}
    \label{eq:own_commit}
\end{equation}
}

\noindent where $\mathrm{\#commits}(d,f)$ is the number of code changes (i.e., commits) that a developer $d$ has contributed to a file $f$ and $\mathrm{\#total\_commits}(f)$ is the total number of code changes made to the file $f$.

\item \textbf{Line-based Code Ownership.} This measure calculates the percentage of code written by a particular developer or team in relation to the total amount of code in the project.
The line-based ownership is used by Rahman and Devanbu~\cite{rahman2013and} to measure the proportion of lines that a developer has contributed to a file relative to the total number of lines for that file.
Formally, the line-based code ownership of a given developer $d$ for a file $f$ is computed as follows:

{\small
\begin{equation}
    \mathrm{OWN\_LINE}(d,f) = \frac{\mathrm{\#lines}(d,f)}{\mathrm{\#total\_lines}(f)}
    \label{eq:own_line}
\end{equation}
}

\noindent where $\mathrm{\#lines}(d, f)$ is the number of lines that a developer $d$ has authored in a file $f$ and $\mathrm{\#total\_lines}(f)$ is the total number of lines of the file $f$.

\end{itemize}

\subsection{Motivating Scenarios}\label{sec:differences}
Code ownership approximation is a useful tool for understanding who is responsible for maintaining and creating a quality improvement plan.
However, the commit-based and line-based code ownership approximations focus on different aspects (contribution vs authorship) which are measured based on different data  (commits vs lines). 
They may identify different sets of developers, leading to inconsistent ownership attribution, and suboptimal software quality improvement plans.
In this section, we conceptualize the problem of code ownership approximations through three motivating scenarios.
Figure \ref{fig:motivating} illustrates each of the motivating scenarios.


\textit{\textbf{Scenario 1: Single-large vs. many-small commits.}} A developer who writes a large proportion of source code in a file in one commit (i.e., one change), can be considered a major developer by the line-based approach but a minor developer by the commit-based approach.
On the other hand, a developer who contributes many commits to a few lines of code can be considered a major developer by the commit-based approach but a minor developer by the line-based approach.
For example, in Figure \ref{fig:motivating} (Scenario 1), 
the commit-based approach will consider \texttt{Chris} as having lower ownership of \texttt{A.java} than \texttt{Pat}, while the line-based approach will consider \texttt{Chris} as having higher ownership than \texttt{Pat}.

\textit{\textbf{Scenario 2: Multiple contributors of the same lines.}} It is possible that a developer made many commits to a file but the changed lines were \emph{overwritten} by another developer.
In this scenario, the first developer who made many commits will be considered a major developer by the commit-based approach.
However, the line-based approach will not be able to capture the contributions of the first developer since the second developer is considered an owner of those overwritten lines.
For example, in Figure \ref{fig:motivating} (Scenario 2), 
the commit-based approach will consider \texttt{Jane} as having higher ownership of \texttt{B.java} than \texttt{Bob}, while the line-based approach will consider \texttt{Bob} as having \textit{higher} ownership than \texttt{Jane}.


\textit{\textbf{Scenario 3: Little changes in a development cycle.}} It is possible that a developer writes many source code lines to a file in a previous release.
However, the developer makes little changes to the file in the current release.
Then, the developer will not be considered an owner of this file by the commit-based approach, while still being considered as a major developer by the line-based approach.
For example, in Figure \ref{fig:motivating} (Scenario 3), 
the commit-based approach will consider \texttt{Mary} as an owner of \texttt{C.java}, while could not identify \texttt{Mary} as a developer of \texttt{C.java} in \texttt{Release 2.0}.
On the other hand, the line-based approach still considers \texttt{Linda} as an owner of \texttt{C.java} as she still owns many lines in the file.

%% file: sections/part-one.tex
\begin{figure*}[t]
    \centering{
    \includegraphics[width=\textwidth]{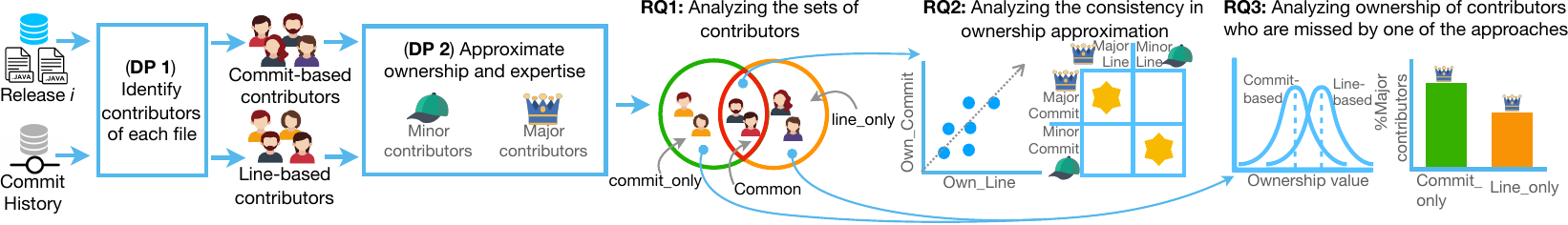}
    \caption{An overview diagram of our data preparation and analysis approaches for RQ1-RQ3.}
    \label{fig:part1-overview}}
\end{figure*}
\section{\analysisI}\label{sec:analysisone}

\subsection{Motivation}
As discussed in Section \ref{sec:differences}, approximating code ownership based on different aspects may lead to different sets of \cnter{s}, different ownership values, and inconsistent expertise levels.
Moreover, it is also possible that some of the major developers (developers who made significant amounts of contributions) could have been missed. 
Hence, we set out to three RQs to examine the consistency in terms of the set of \cnter{s} identified by the two approaches (RQ1), ownership values and expertise levels (RQ2), and the number of major developers that may not be identified (RQ3).
Figure \ref{fig:part1-overview} provides an overview of our data preparation and analysis approaches.

\input{sections/table_studied_systems.tex}
\subsection{Studied Systems}
In this work, we use a corpus of publicly-available defect datasets provided by Yatish~\ea~\cite{yatish2019mining} where the ground truths were labelled based on the affected releases. 
We opt to use these datasets because prior work~\cite{yatish2019mining}  showed that when using the approximation of a post-release window (e.g., 6 months) to identify post-release defects, (1) some issue reports addressed within the post-release window do not affect a studied release (false positive), while (2) some issue reports that affect the studied release can be addressed later than the post-release window (false negative). 

For each system, the top three releases with the highest number of issue reports that affected the release were selected. This is to ensure that defect prediction models are trained on high-quality datasets with sufficient positive samples of data (i.e., defective files). 
Thus, we conduct a study based on 25 releases span across seven open-source software systems. 
Each dataset has 59 software metrics along two
dimensions, i.e., 54 product metrics and 5 process metrics. 
Table~\ref{studied_datasets} presents a summary of the studied systems.

\subsection{Data Preparation}
\input{sections/data_preparation}

\subsection*{\textbf{RQ1: \rqone}}

\smallsection{Approach}
To answer our RQ1, we analyze the sets of developers identified by the commit-based and line-based approaches for each file in the studied releases.
More specifically, we first examine how many developers can be identified by both approaches, i.e., $\mathrm{common} = \frac{|D_{C} \cap D_{L}|}{|D_{C} \cup D_{L}|}$, where $D_C$ is a set of developers identified by the commit-based approach and $D_L$ is a set of developers identified by the line-based approach.
Then, we examine a proportion of developers who can only be identified by the commit-based approach (i.e., $\mathrm{commit\_{only}} = \frac{|D_{C} - D_{L}|}{|D_{C} \cup D_{L}|}$) and those who can only be identified by the line-based approach (i.e., $\mathrm{line\_{only}} = \frac{|D_{L} - D_{C}|}{|D_{C} \cup D_{L}|}$).

\begin{figure}[t]
    \includegraphics[width=\columnwidth]{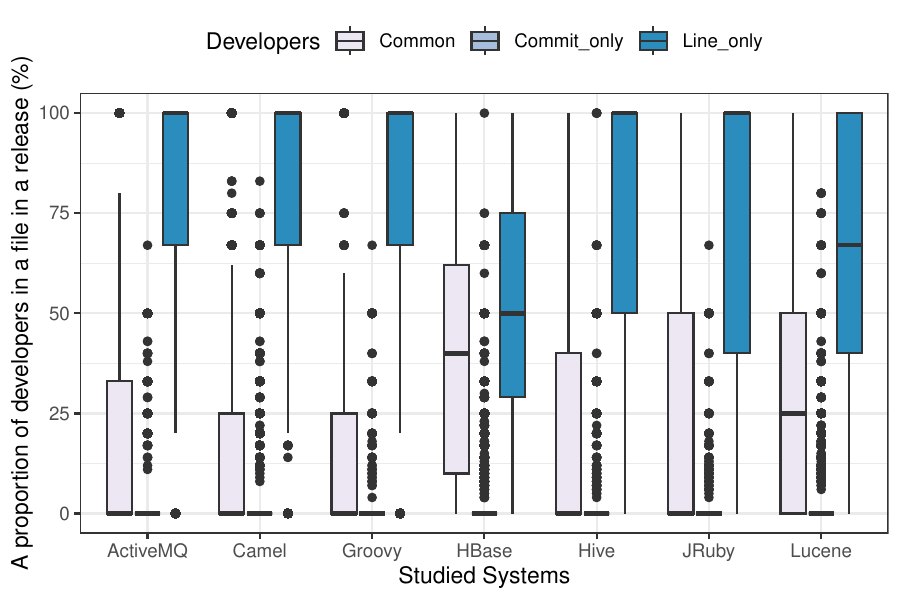}
    \caption{A proportion of developers that are identified by both approaches (common), by only the commit-based approach (commit\_only), and by only the line-based approach (line\_only).}
    \label{fig:prop_authors}
\end{figure}

\smallsection{Results} \textbf{Only 0\% to 40\% of developers in a file can be commonly identified by commit-based and line-based approaches.}
Figure \ref{fig:prop_authors} shows that at the median value, the proportions of common developers is 0\% for ActiveMQ, Camel, Groovy, Hive, and JRuby.
We find that 25\% (HBase) to 70\% (Camel) of the files do not have common developers who can be identified by both approaches.
Only Lucene and HBase have a median value of 25\% and 40\% of developers that can be commonly identified by both approaches, respectively.
Nevertheless, none of the files in our studied systems have a proportion of common developers of 100\%.
These results suggest that the commit-based and line-based approaches do not identify the same set of developers for a file.

Moreover, Figure \ref{fig:prop_authors} shows that a substantial proportion of developers can be missed when using the commit-based approach.
At the median value, the proportion of line\_only developers ranges from 10\% to 100\% of the total developers in a file, indicating that the commit-based approach could overlook a substantial proportion of developers that can be identified by the line-based approach.
As discussed in Section \ref{sec:differences}, this inconsistency in the sets of developers identified by the two approaches may be because of little changes in a development cycle (Scenario 3).
We observe that 19\% (HBase) to 67\% (Camel) of the files were not changed during a development cycle.\footnote{Note that although these files were not changed, 0\% - 18\% (with a median of 2\%) of these files have a post-release defect.}
This suggests that if the files were not changed during the development cycle, the commit-based approach will not be able to identify any developers of those files.

\begin{tcolorbox}[title= \textbf{RQ1 Summary:}, left=2pt, right=2pt, top=2pt,bottom=2pt]
Only 0\% to 40\% of developers can be identified by both commit-based and line-based approaches, indicating these approaches do not identify the same set of developers. A substantial proportion of developers can be missed when using one of the ownership approximation approaches. 
\end{tcolorbox}

\subsection*{\textbf{RQ2: \rqtwo}}
\smallsection{Approach}
To address our RQ2, we analyze the code ownership values of the common developers who can be identified by both commit-based and line-based approaches.
To do so, for each file in the studied releases, we analyze the correlation between the ownership values of each developer approximated by the commit-based and line-based approaches.
We use Spearman's correlation analysis.
A correlation coefficient $|\rho|$ value of 1 indicates that the commit-based and line-based approaches approximate ownership values in a consistent direction (e.g., the more the commit-based ownership value, the more the line-based ownership value).
On the other hand, a $|\rho|$ value of 0 indicates that the commit-based and line-based approaches do not consistently approximate the ownership values.
The correlation is considered as weak for $|\rho| < 0.3$, moderate for $ 0.3 \leq |\rho| < 0.7 $ , and strong for $|\rho| \geq 0.7$~\cite{dancey2007statistics,akoglu2018user}.



\begin{figure}[t]
    \includegraphics[width=\columnwidth]{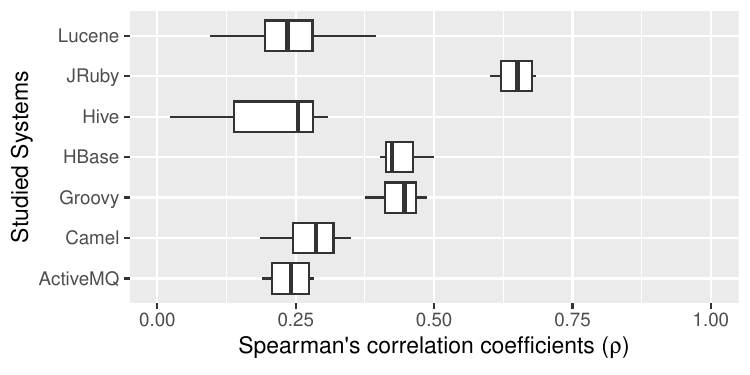}
    \caption{The Spearman's correlation coefficient between the ownership values of the commit-based and line-based approaches.}
    \label{fig:own_correlation}
\end{figure}

\begin{figure*}[t]
    \includegraphics[width=1.0\textwidth,trim=0 20 0 0, clip]{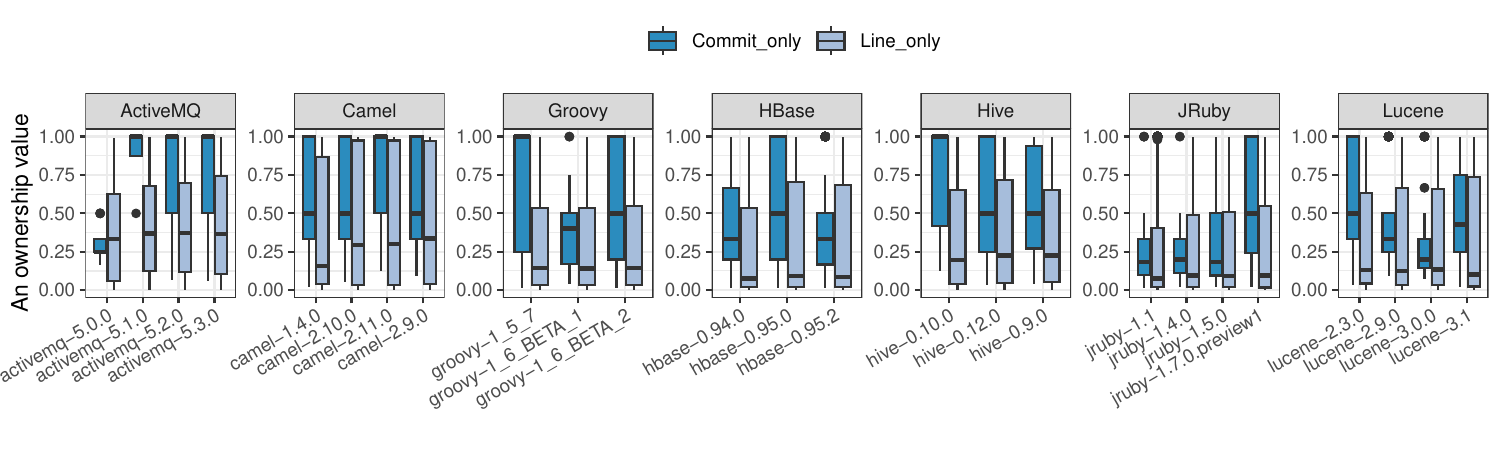}
    \caption{A distribution of an ownership value for the \texttt{commit\_only} and \texttt{line\_only} developers.}
    \label{fig:ownership_distribution}
\end{figure*}
\smallsection{Results}
\textbf{The ownership values approximated by the commit-based and line-based approaches have a small to moderate correlation.}
Figure \ref{fig:own_correlation} shows the distribution of Spearman's correlation coefficient ($\rho$) of the ownership values approximated by the commit-based and line-based approaches.
For four out of the seven studied systems, the correlation is small with the median $\rho$ value of 0.24 (ActiveMQ, Lucene) - 0.29 (Camel).
For the other three systems (i.e., Groovy, HBase, and JRuby), the correlation is moderate with the median $\rho$ value of 0.42 (HBase) - 0.65 (JRuby).
These results indicate that the ownership values approximated by the commit-based and line-based approaches are weakly to moderately correlated.
As discussed in Section \ref{sec:differences}, this weak correlation of ownership values could be related to the single-large vs many-small commits or multiple contributions of the same lines (Scenario 1 \& 2).


Nevertheless, when we examine the expertise levels of these common developers, we find that a majority of them are classified into a consistent expertise level.
In particular, we find that on average across studied releases, 56\% (HBase) to 80\% (Groovy) of the common developers have a consistent expertise level.
In particular, these common developers are consistently classified as major developers, while few of the common developers (0\%-4\%) are consistently classified as minor developers.
These results indicate that the commit-based and line-based approaches consistently identify the expertise level of the majority of the common developers.



\begin{tcolorbox}[title= \textbf{RQ2 Summary:}, left=2pt, right=2pt, top=2pt,bottom=2pt]
For the developers who can be commonly identified by both commit-based and line-based approaches, their ownership values are different and not highly consistent. Nevertheless, a majority (50\% to 93\%) of these developers still have a consistent expertise level by the two approaches.

\end{tcolorbox}



\subsection*{\textbf{RQ3: \rqthree}}
\smallsection{Approach} Our RQ1 shows that a large proportion of developers can be missed when using one of the approximation approaches.
We set out to further investigate the ownership values of these developers.
We check which group of the developers (commit\_only or line\_only) made a larger contribution.
We use a one-sided Wilcoxon rank sum test to determine whether the ownership values of the commit\_only developers are statistically greater (or less) than the ownership values of the line\_only developers.
We also use Cliff's $|\delta|$ to measure the effect size (i.e., the magnitude of the difference).
The difference is considered as negligible for $|\delta| < 0.147$, small for $0.147 \leq |\delta| < 0.33$, medium for $0.33 \leq |\delta| < 0.474$, and large: $0.474 \leq |\delta|$.


\smallsection{Results}
\textbf{The commit\_only developers typically have an ownership value of 0.18-1, while the line\_only developers typically have an ownership value of 0.07-0.37.}
Figure~\ref{fig:ownership_distribution} shows a distribution of ownership values for the commit\_only and line\_only developers for each studied release.
At the median value, the ownership values of the commit\_only developers range from 0.18 (JRuby) - 1 (ActiveMQ, Camel, Groovy, and Hive).
On the other hand, the ownership values of the \texttt{line\_only} developer range from 0.07 (JRuby) - 0.37 (ActiveMQ).
The statistical analysis also confirms that the ownership values of the commit\_only developers are statistically larger than the ownership values of the line\_only developers ($p$-value $<$ 0.05) with a small to large effect size ($|\delta|$ is 0.17 - 0.74)  for all the studied releases (except ActiveMQ 5.0.0).
These findings suggest that the line-based approach misses the commit\_only developers who made a relatively large contribution (in terms of the number of commits) during the development cycle.

When we examine the expertise levels, we find that a large proportion of  commit\_only and line\_only developers are major developers.
In particular, on average across studied releases in each system, 89\% (JRuby) - 100\% (ActiveMQ, Camel, Hive, and Lucene) of the commit\_only developers are considered major based on the number of commits made to a file.
On the other hand, 67\% (HBase) - 86\% (ActiveMQ) of the line\_only developers are considered major developers based on the number of lines authored in a file in the released version.
These results suggest that a large proportion of the developers who are missed by one of the ownership approximation approaches are major developers.

\begin{tcolorbox}[title= \textbf{RQ3 Summary:}, left=2pt, right=2pt, top=2pt,bottom=2pt]
The developers who are missed by the line-based approach (i.e., commit\_only developers) have an ownership value that is statistically higher than the developers who are missed by the commit-based approach (i.e., line\_only developers).
Moreover, a large proportion of the commit\_only and line\_only developers are major developers, suggesting that major developers can be missed when using one of the ownership approximation approaches.
\end{tcolorbox}



%% file: sections/table_studied_systems.tex
\begin{table*}[t]
\caption{A statistical summary of the studied systems.}
\label{studied_datasets}
\centering
\resizebox{\textwidth}{!}{
\begin{tabular}{llllll} 
\hline
Name & Description & No. of files & Defective Rate & KLOC & Studied Releases \\
\hline
ActiveMQ & Messaging and Integration Patterns server & 1,884-3,420 & 6\%-15\% & 142-299 &5.0.0, 5.1.0, 5.2.0, 5.3.0 \\ 
Camel & Enterprise Integration Framework & 1,515-8,846 & 2\%-18\% & 75-383 & 1.4.0, 2.9.0, 2.10.0, 2.11.0 \\
Groovy & Java-syntax-compatible OOP for JAVA & 757-884 & 3\%-8\% & 74-90 & 1.5.7, 1.6.0.Beta\_1, 1.6.0.Beta\_2\\
HBase & Distributed Scalable Data Store & 1,059-1,834 & 20\%-26\% & 246-534 &  0.94.0, 0.95.0, 0.95.2 \\
Hive & Data Warehouse System for Hadoop & 1,416-2,662 & 8\%-19\% & 287-563 & 0.9.0, 0.10.0, 0.12.0 \\
JRuby & Ruby Programming Lang for JVM & 731-1,614 & 5\%-18\% & 105-238 & 1.1, 1.4, 1.5, 1.7\\ 
Lucene & Text Search Engine Library & 805-2,806 & 3\%-24\% & 101-342 & 2.3.0, 2.9.0, 3.0.0, 3.1.0 \\
\hline
\end{tabular}
}
\end{table*}

%% file: sections/data_preparation.tex

\subsubsection*{(Step 1) Identify developers of each file} 

\underline{Commit-based Approach}: To obtain a set of commit-based contributors, we first retrieve a list of commits that (1) occurred on the release branch or (2) that
originated on other branches, but have been merged into the release branch.
Similar to prior work~\cite{Bird2011a,bird2010analysis,rahman2013and}, since we aim to investigate the association between the code ownership of a file and the likelihood that a file will have a post-release defect, we consider only the commits that were merged into the branch of a studied release to measure development activities during the development cycle of that release.
Then, we use \texttt{git name-rev} to identify the commits that are associated with a studied release.
For each commit, we retrieve the name and email address of the contributor and the list of changed files.
Finally, we count the number of commits that a commit-based contributor made to a file.

\underline{Line-based Approach}: To obtain a set of line-based contributors, we first retrieve a list of files at the released version. 
Then, for each file, we use the \texttt{git blame} command to identify a contributor who recently modified each line in the file.
Finally, we count the number of lines that a line-based contributor modified in the file. 

\subsubsection*{(Step 2) Compute ownership values and identify their levels of expertise.} 
We compute ownership values as described in Section \ref{sec:background}. 
We apply these two calculations to each developer of each file.
Then, we identify the expertise level of a developer.
Following Bird~\ea~\cite{Bird2011a,bird2010analysis}, the expertise level can be defined based on the code ownership values.  
Developers with a high code ownership value (e.g., above 5\%) are considered as \textbf{\textit{major}} developers, while developers with a low code ownership value (e.g., below 5\%) are considered as \textbf{\textit{minor}} developers.

%% file: sections/part-two.tex
\section{\analysisII} \label{sec:analysistwo}

\begin{figure*}[t]
\centering
\includegraphics[width=.7\textwidth]{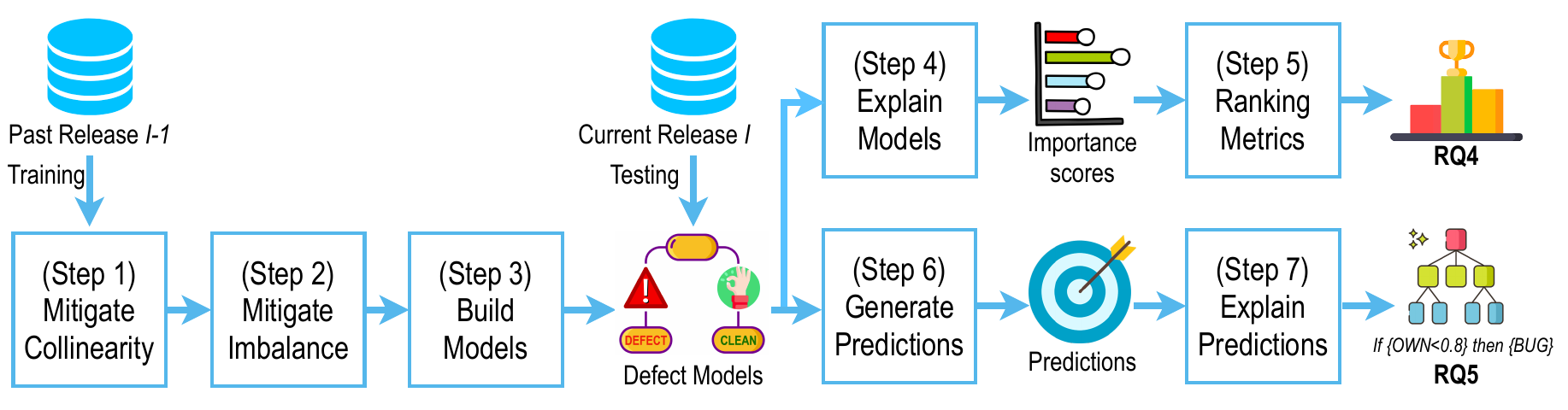}
\caption{An overview diagram of our approach for analyzing the association of code ownership approximations and software quality.}
\label{fig:part2-overview}
\end{figure*}

\subsection{Motivation}


In practice, code ownership is widely used to identify who is responsible for bug fixing and improving code quality.
Prior studies suggested building a defect prediction model to understand the association between code ownership approximations and software quality~\cite{Bird2011a,thongtanunam2016revisiting}.
However, the results of Section \ref{sec:analysisone} show that the commit-based and line-based ownership approximations are different in terms of a set of developers and ownership values.
However, it remains unclear about which code ownership approximation approaches have the strongest association with software quality.
Thus, we set out to investigate which code ownership measures are the most important, and whether code ownership that is important in a past release is also important in a subsequent release.





\subsection{Model Building}

We developed defect prediction models to investigate the association between code ownership approximations and software quality.
We build a defect prediction model in a cross-release setting (i.e., releases $i-1$, $i$).
Borrowing the concept of Explainable AI, we focus on both explaining the model (\emph{global explanation}) trained from the past release $i-1$ and explaining the predictions (\emph{local explanations}) of the defective files in the current release $i$.
The explanation is presented in the form of the most important metrics.
We consider the following features in our model.
Figure \ref{fig:part2-overview} presents an overview of our model-building approach.
Below, we describe the studied metrics, the confounding metrics, and the model-building steps in detail.






%






\smallsection{Studied Metrics:} We studied the six ownership metrics, i.e., two ownership values (i.e., OWN\_COMMIT and OWN\_LINE) approximated by the commit-based and line approaches as described in Section \ref{sec:background} and the number of developers in each of four expertise levels (i.e., MAJOR\_COMMIT, MINOR\_COMMIT, MAJOR\_LINE and MINOR\_LINE).

\smallsection{Confounding Metrics}
Prior studies have shown that several software factors can have an impact on software quality~\cite{Hassan2009, yatish2019mining, rahman2011ownership, hall2012systematic}.
Since we want to analyze the association between code ownership approximations and software quality, we use the 54 product and five process metrics provided by \citet{yatish2019mining} as confounding factors.
\textit{Product metrics} measure code characteristics in three dimensions: (1) complexity, e.g., McCabe Cyclomatic, (2) volume, e.g., lines of code, and (3) object-oriented design, e.g., coupling between object classes.
\textit{Process metrics} measure the development activity during the development cycle of a release, e.g., the number of commits, and the number of lines added and deleted.

\textit{(Step 1) Mitigate  Multi-collinearity.} 
We first analyze and remove multi-collinearity among studied metrics and confounding metrics to avoid the spurious relationship of the metrics and defect-proneness~\cite{jiarpakdee2019collinearity}. 
We used the AutoSpearman approach~\cite{jiarpakdee2018icsme}, which is an automated approach based on the Spearman correlation analysis and the Variance Inflation Factor analysis (VIF) to remove multi-collinear metrics.
\textit{(Step 2) Mitigate Class Imbalance.} 
Class imbalance is a phenomenon of a defect dataset where the proportion of defective and clean modules is not equally represented which can have a negative impact on the predictions of defect prediction models~\cite{tantithamthavorn2019imbalance}.
Thus, we apply the SMOTE technique to each of the training datasets using the \texttt{smote} function of the \texttt{imblearn} python package.

\textit{(Step 3) Training a Model.} 
We train models based on the cross-release scenario, i.e., a previous release $i-1$ is used for training, and a current release $i$ is used for evaluation.
We use the Random Forests classification technique since the Random Forests classification technique tends to produce the most accurate defect prediction models ~\cite{ghotra2015revisiting,tantithamthavorn2016automated,tantithamthavorn2019optimization}.
We build the model using the \texttt{RandomForestClassifier} of the \texttt{scikit-learn} package.
In addition, as suggested by prior work~\cite{tantithamthavorn2016automated,tantithamthavorn2019optimization}, we optimize hyper-parameter settings of the random forest classification technique using the \texttt{RandomizedSearchCV} function. 

\begin{figure}[t]
\centering
\includegraphics[width=0.8\columnwidth, trim = 0 5 0 0, clip = true]{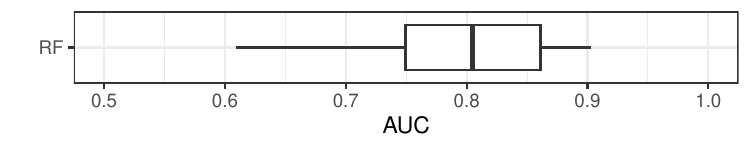}
\caption{The AUC values of our defect prediction models.}
\label{fig:AUC}
\end{figure}

Finally, we evaluate the accuracy of the models based on the data of the current release $i$ using the Area Under the receiver operator characteristic Curve (AUC). The AUC measure is a threshold-independent performance measure that evaluates the ability of models to discriminate between defective and clean files, where the AUC value ranges between 0 (worst), 0.5 (no better than random guessing), and 1 (best)~\cite{Hanley1982}.
Figure~\ref{fig:AUC} shows that our defect prediction models achieve a median AUC value of 0.80, confirming that the models are accurate enough to be used for further analysis.



\begin{figure}[t]
\includegraphics[width=0.65\columnwidth, trim={0 1cm 18.8cm 0},clip, angle =-90]{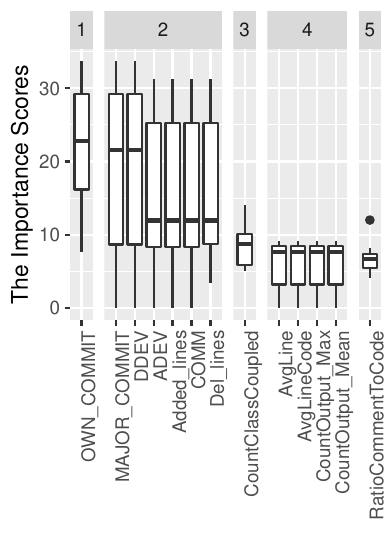}
\caption{(RQ4) The top-5 non-parametric ScottKnott ESD (NPSK) ranking of the feature importance scores of our defect prediction models.}
\label{fig:rq4-varimp}
\end{figure}

\begin{figure*}[t]
\centering
\includegraphics[width=.7\textwidth, trim = 0 23 0 0, clip=true]{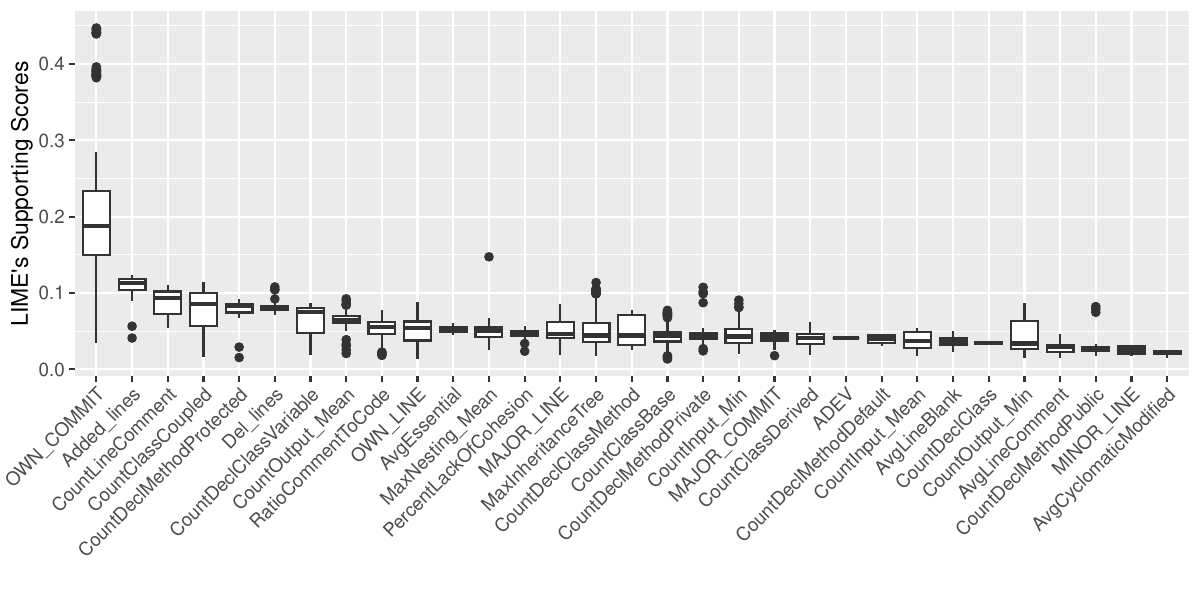}
\caption{(RQ5) The LIME's supporting scores of each metric when explaining the prediction of defective files in the current release.}
\label{fig:rq5-lime}
\end{figure*}

\begin{figure}[t]
\includegraphics[width=1.0\columnwidth]{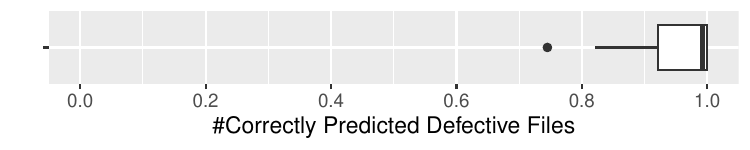}
\caption{(RQ5) The proportion of correctly predicted defective files that code ownership metrics appear at the top rank of LIME's explanations.
}
\label{fig:rq5-result}
\end{figure}

\subsection*{\textbf{RQ4: \rqfour}} 

\noindent \smallsection{Approach} 
We perform the following two steps.



\textit{(Step 4) Explain the Defect Prediction Models.}
To analyze the importance of code ownership metrics, we analyze the defect prediction models by computing the metric importance scores of defect prediction models.
We use the calculation of the permutation importance that is built within the \texttt{feature\_importances\_} of the random forest technique in the \texttt{scikit-learn} Python package.




\textit{(Step 5) Ranking the Most Important Metrics.}
Since the distributions of importance scores are estimated based on multiple models of multiple releases, we use the Non-Parametric ScottKnott ESD (NPSK) test\footnote{https://github.com/klainfo/ScottKnottESD} to find statistically distinct ranks of metrics based on the importance scores.
The NPSK test is a multiple comparison approach that leverages hierarchical clustering to partition the set of median values of techniques (e.g., medians of variable importance scores, medians of model performance) into statistically distinct groups with a non-negligible difference.
Unlike the original ScottKnott ESD test~\cite{tantithamthavorn2017empirical}, the NPSK test does not require the assumptions of normal distributions, homogeneous distributions, and the minimum sample size.

\smallsection{Results} 
\textbf{Commit-based code ownership approximation has the highest important scores.}
Figure~\ref{fig:rq4-varimp} shows the non-parametric ScottKnott ESD ranking of the importance scores of the studied metrics in the software defect models.
The ownership values approximated by the commit-based approach (OWN\_COMMIT) has the highest rank of the important scores in our defect models. 
Moreover, the number of major developers based on commit-based ownership (MAJOR\_COMMIT) also has the second highest rank of the importance scores.
On the other hand, we find that the metrics of the line-based ownership approximation (e.g., OWN\_LINE) appear to have the sixth rank of the important scores.

\begin{tcolorbox}[title= \textbf{RQ4 Summary:}, left=2pt, right=2pt, top=2pt,bottom=2pt]
Considering all ownership metrics and confounding factors, the commit-based code ownership approximation has the highest important scores to our software defect models, suggesting that the commit-based code ownership metrics share the strongest relationship with software quality.
\end{tcolorbox}



\subsection*{\textbf{RQ5: \rqfive}}

\smallsection{Approach} Since the importance scores in RQ4 are derived from the models that are trained from the data of a previous release ($i-1$), we further investigate whether the importance scores still apply to all predictions in the current release $i$ of the same project. 
To do so, we perform the following step.

\textit{(Steps 6,7) Generate and Explain the Predictions of Defect Models.}
For each release $i$, we use our defect model trained based on the previous release $i-1$ data to predict whether each file will be defective.
Then, we use a state-of-the-art Local-Interpretability Model-agnostic Explanations (LIME) technique to explain each prediction of the defect models~\cite{ribeiro2016should}. 
LIME is one of the most well-known local model-agnostic techniques for explaining an individual prediction of any classification technique.
We use the implementation of the LIME technique that is provided by the \texttt{lime} python package.
Given a prediction, an explanation of LIME includes a supporting score to indicate the importance of a metric that the defect model uses for the prediction.

To address RQ5, we analyze the supporting scores of the top 5 metrics for the correctly predicted defective files.
Specifically, we examine the distributions of LIME's supporting scores of the top 5 metrics.
In addition, we measure the proportion of the correctly predicted defective files whose code ownership metrics have the highest supporting score.


\smallsection{Results} \textbf{When predicting defective files in the current release, the commit-based code ownership metric has a higher supporting score in predictions than other metrics.}
Figure~\ref{fig:rq5-lime} shows the distribution of supporting scores of the top 5 metrics in predictions of files in the current release. 
We found that the ownership value approximated by the commit-based approach (i.e., OWN\_COMMIT) has supporting scores in predictions significantly higher than other metrics. 
The supporting score varies from 0.04 to 0.45 with a median of 0.18, while other metrics vary from 0.01 to 0.15 with a median of 0.05.
The one-sided Wilcoxon Rank Sum tests also confirm that the supporting scores of OWN\_COMMIT are statistically higher than each of the other metrics ($p$-value $< 0.001$) with a large Cliff's delta effect size difference.
These results indicate that the commit-based ownership metric is often the most important metric for predicting defective files.




\textbf{The commit-based code ownership metric typically has the highest supporting scores in prediction for 97\%(median) of the correctly predicted defective files.}
Figure~\ref{fig:rq5-result} shows the percentage of the correctly predicted defective files for the current releases whose OWN\_COMMIT has the highest LIME's supporting score.
We found that at the median value, 97\% of the files in a release have OWN\_COMMIT with the highest supporting score.
Specifically, in five studied releases, all of the correctly predicted defective files have OWN\_COMMIT as the highest supporting scores
(i.e, activemq-5.1.0, camel-2.9.0, camel-2.10.0, groovy-1\_6\_BETA\_1, lucene-2.9.0).
For the other studied releases, the proportion of the files where OWN\_COMMIT has the highest supporting score varies between 75\% to 97\%.
These results suggest that when predicting defective files in the current release, the commit-based code ownership metric is still the most important metric for predictions.


\begin{tcolorbox}[title= \textbf{RQ5 Summary:}, left=2pt, right=2pt, top=2pt,bottom=2pt]


The commit-based code ownership which is important in the past release is also important in the current release, suggesting that commit-based code ownership should be used to explain why a file is predicted as defective and guide the development of QA improvement plans.
\end{tcolorbox}

%% file: sections/relatedwork.tex
\section{Related Work}\label{sec:relatedwork}


\subsection{Code Ownership Empirical Studies}

While many studies adopted code ownership approximations to investigate the impact on code ownership in several aspects~\cite{borg2023u,anvik2006should,linares2012triaging,borg2023u,thongtanunam2017review}, few studies empirically investigate the differences in these metrics.
\citet{avelino2019Who} and \citet{Cury2022Identifying} investigated the accuracy of three ownership approaches to identify developers who declared themselves as knowledgeable developers for a file.
\citet{Hannebauer2016Automatically} compare the performance of code ownership approximations for code reviewer recommendation. 
\citet{thongtanunam2016revisiting} investigate the differences between commit-based ownership and review-based ownership.
\citet{oliveira2020code} investigate the team leaders' perception of different code ownership metrics for measuring productivity.

\smallsection{Key Difference} While prior studies compare the performance of code ownership metrics for various specific tasks, none of the studies empirically quantify the difference in the set of identified developers, and the code ownership values and the association of ownership metrics with software quality.

\subsection{Code Ownership \& Software Quality}

There is a plethora of studies that investigate the relationship between code ownership and software quality.
For example, Seifert~\ea~\cite{illes2008exploring} found that an increasing number of distinct authors making changes to a file may lead to more software defects. 
Meneely~\ea~\cite{meneely2009secure} found that contributions by less-focused developers were associated with more security-related errors. 
A study of Bird~\ea~\cite{bird2010analysis} at Microsoft products found that small commit-based contributions by minor developers are associated with the number of post-release defects.
A study by Rahman and Devanbu~\cite{rahman2011ownership} found that line-based code ownership has a large impact on software quality.
In contrast, Weyuker~\ea~\cite{weyuker2008too} and Graves~\ea~\cite{graves2000predicting} found that the number of contributors does not have a strong relationship with defect-proneness.

\smallsection{Key Difference} Since the conclusions of prior studies are contradictory, we suspect that the different conclusions may have to do with the use of different code ownership approximations (e.g., commit-based or line-based code ownership).
However, there exist no studies that investigate the impact of both commit-based and line-based code ownership approximations on software quality.
In this paper, we address this challenge by investigating the nature of the differences between the commit-based and line-based code ownership approximations and their impact on software quality.

\subsection{Code Ownership for Software Quality Improvement Plans}

The findings of prior work draw several important implications for practitioners to develop software quality improvement plans based on code ownership.
For example, a study at Microsoft~\cite{Bird2011a,bird2010analysis} recommended that files with low code ownership should be given priority by software quality assurance resources (i.e., more testing and code review effort).
Another study at Microsoft~\cite{nagappan2008influence,bird2009does} also found that when more people work on software components, it has more failures since there is no clear point of contact.
Thus, the contributions to a software component are spread across many developers, and there is an increased chance of communication breakdowns, misaligned goals, inconsistent interfaces, and semantics, all leading to poor software quality.

Rahman and Devanbu~\cite{rahman2011ownership} commented that these implications are very crucial for software-intensive organizations to develop the most effective quality improvement plans.
Thus, managers and team leads can make better decisions about how to govern a project.
If code ownership has the largest impact on software quality, then software quality improvement plans to enforce strong code ownership can be put into place.
On the other hand, if code ownership has little effect, then the normal bottlenecks associated with having one person in charge of each component can be removed, and available talent reassigned at will.

\smallsection{Key Difference} While prior studies only derive the importance of code ownership on software quality based on a past release, there exist no studies investigating if the importance of such metrics in the past still holds in the current release and would the risk of having defects be decreased when software quality improvement plans to enforce strong code ownership is put into place.
